\documentclass[aps,floatfix,twocolumn,prl]{revtex4}
\usepackage{graphicx,subfigure,amsmath,bbm}
\pdfoutput=1

\newcommand{\ua}{\uparrow}
\newcommand{\da}{\downarrow}

\def\braket#1{\mathinner{\langle{#1}\rangle}}
\def\bra#1{\left\langle#1\right|}
\def\ket#1{\left|#1\right\rangle}

\begin{document}

\title{Multiple Nuclear Polarization States in a Double Quantum Dot}
\date{\today}

\author{J. Danon}
\author{I.T. Vink}
\author{F.H.L. Koppens}
\author{K.C. Nowack}
\author{L.M.K. Vandersypen}
\author{Yu.V. Nazarov}
\affiliation{Kavli Institute of NanoScience, Delft University of Technology, 2628 CJ Delft, The Netherlands}

%\pacs{??}
%{03.67.Lx,73.63.Kv,76.30.-v}
%73.23.Hk Coulomb blockade; single-electron tunneling
%03.67.Lx Quantum computation
%03.67.-a Quantum information
%76.30.-v Electron paramagnetic resonance and relaxation
%73.63.Kv Quantum dots

\begin{abstract}
We observe multiple stable states of nuclear polarization and nuclear self-tuning over a large range of fields in a double quantum dot under conditions of electron spin resonance. The observations can be understood within an elaborated theoretical rate equation model for the polarization in each of the dots, in the limit of strong driving. This model also captures unusual features of the data, such as fast switching and a `wrong' sign of polarization. The results reported enable applications of this polarization effect, including accurate manipulation and control of nuclear fields.\end{abstract}

\maketitle

%\section{Introduction}

%Due to their weak coupling to the environment, electron spins confined in semiconductor quantum dots are an attractive candidate to implement \emph{qubits}, i.e.\ the computational units in a quantum computer~\cite{PhysRevA.57.120}.
Great experimental progress in the last decade enabled the confinement, initialization and read-out of single spins in quantum dots~\cite{read-out}. Controlled coherent single-spin rotations --- a key ingredient for quantum manipulation --- were demonstrated recently using the electron spin resonance (ESR)~\cite{frank:nature,katja:science,laird:246601,kroner:156803,pioro:nature}. The weak hyperfine coupling of the electron spin to the nuclear spins in the host material appeared to be of great importance in this field. It was identified as the main source of qubit decoherence and provides a significant hybridization of the spin states~\cite{frank:science,HFdecoherence}. This has stimulated intensive theoretical and experimental research focusing on nuclear spin dynamics in quantum dots~\cite{A.C.Johnson:nature,klauser:205302,tartakovskii:026806,maletinsky:056804,D.J.Reilly08082008,korenev:prl}.

Overhauser pointed out already in the the 1950s~\cite{overhauser} that ESR may provide the buildup of significant nuclear spin polarization. Indeed, most ESR experiments on quantum dots, aimed at demonstrating electron spin rotations, also clearly demonstrated dynamical nuclear spin polarization (DNSP)~\cite{frank:nature,katja:science,laird:246601,kroner:156803}. 

For ESR driving of a single spin in an almost isolated quantum dot, or an ensemble of such dots, the scenario is similar to that of the usual Overhauser effect: the direction of DNSP is parallel to the spin of the excited electrons~\cite{overhauser,danon:056603}. Recent ESR experiments on self-assembled quantum dots have confirmed this picture~\cite{kroner:156803}, and a similar reasoning holds for spin experiments with optically pumped dots~\cite{greilich:science}. In some cases, a bistability has been observed: Under the same conditions, the nuclear spins in the dot can be either polarized or unpolarized~\cite{tartakovskii:026806}.

However, several issues can complicate the situation. In recent ESR experiments in double quantum dots~\cite{frank:nature,katja:science,laird:246601} (i) electrons participate in \emph{transport} during ESR driving, and (ii) there can be different nuclear spin dynamics in the two dots.
Furthermore, a driving magnetic field is in practice accompanied by an electric field which modulates the electron-nuclear spin coupling at the resonance frequency~\cite{laird:246601}. All this makes a straightforward extension of existing models~\cite{danon:056603} impossible and promises richer and more interesting physics, which we indeed reveal.

In this Letter, we report a study of ESR in a double quantum dot focusing on DNSP. We have observed multiple stable states of nuclear polarization (up to four states), not seen in single-dot experiments, nuclear self-tuning to the ESR condition over a large range of magnetic fields ($\gtrsim$ 100 mT), and a sign of DNSP opposite to that following from the Overhauser argument. We identify the most probable mechanism governing DNSP and present a theoretical model explaining our findings. The results reported enable applications of this self-tuning effect, including accurate manipulation and control of the nuclear polarization~\cite{ivo:pc} and use of this for improving the electron spin coherence time, possibly by orders of magnitude.

\begin{figure}[b]
\includegraphics[width=8.5cm]{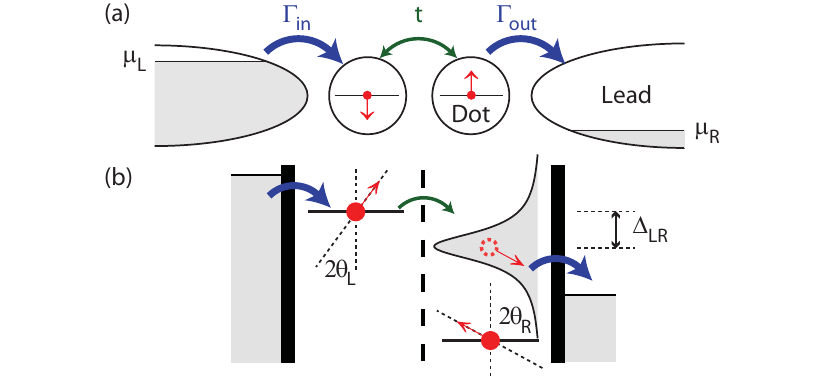}
\caption{Double dot setup. (a) The double quantum dot is coupled to two leads. Due to a voltage bias, electrons can only run from the left to the right lead, implementing the transport sequence $(1,1) \to (0,2) \to (0,1) \to (1,1)$. (b) Energy diagram. The four possible $(1,1)$ states differ in spin projections on the quantization axes (red arrows). Under ESR conditions the axes can be different in the two dots and do not coincide with the direction of the external magnetic field. These states are coherently coupled (green arrow) to the $(0,2)$ singlet that decays quickly (broadened line), leaving the system in $(0,1)$.} \label{fig:fig1}
\end{figure}

\begin{figure*}[t]
\includegraphics[width=17cm]{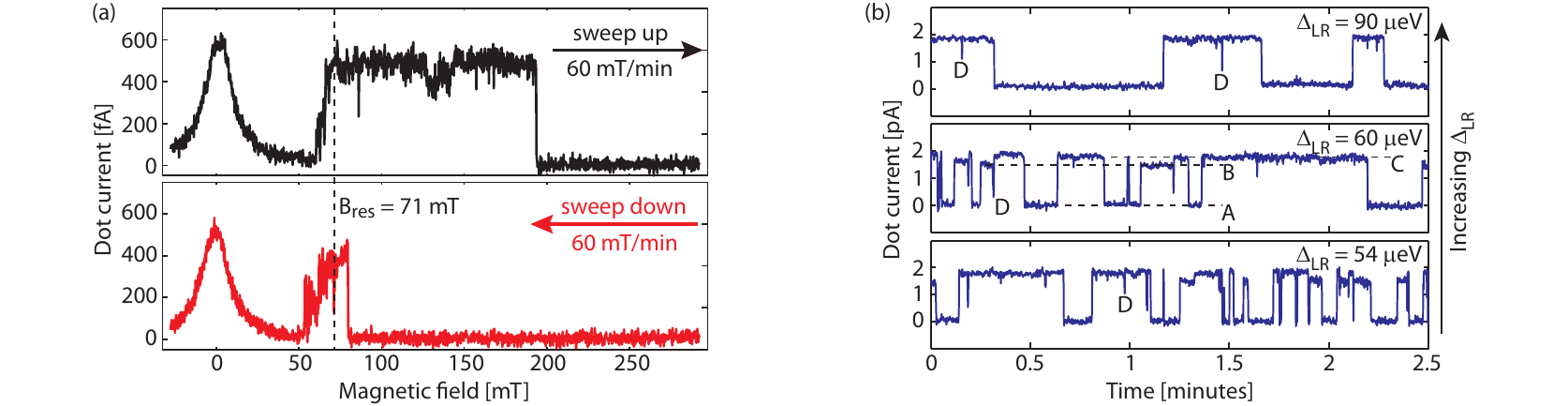}
\caption{(a) Magnetic field sweeps for $\omega$ fixed at 350 MHz. Upper panel: Magnetic field sweep from low to high values resulting in an ESR peak width exceeding 100 mT. Lower panel: Sweep in the opposite direction, showing a much narrower ESR peak~\cite{fig2}. The nominal resonance condition $B_\text{res} = \hbar\omega / g\mu_B$ is met at $B_0 \approx 71$~mT for $\omega = 350$~MHz and $g = 0.35$~\cite{frank:nature} (see dashed line). Note that in both traces the nuclear bath is unpolarized at the onset of electron spin resonance~\cite{supp}. (b) Multiple values of the current through the double dot approximately at resonance. The current switches between at least three stable values on a time scale of seconds to minutes. The three panels correspond to three different values of the energy level detuning $\Delta_{LR}$ (increasing from the bottom to the upper panel). The values given for $\Delta_{LR}$ may have a constant offset, as photon assisted tunneling processes broaden the interdot transition which makes it difficult to separate resonant and inelastic transport. In both (a) and (b) the lowest value of current was subtracted as offset. The data in (b) were taken for a larger $\Gamma_\text{in}$ and $\Gamma_\text{out}$ than the data in (a).} \label{fig:exp}
\end{figure*}

%\section{Experiment}

The double quantum dot system is electrostatically defined in a two-dimensional electron gas, located 90 nm below the surface of a GaAs/AlGaAs heterostructure, by applying negative voltages to metal surface gates. The dots are tuned to the Pauli spin blockade regime~\cite{ono:science}, where the transport sequence of charge states is $(1,1) \to (0,2) \to (0,1) \to (1,1)$, $(n,m)$ denoting the charge state with $n(m)$ excess electrons in the left(right) dot. The current through the double dot depends on the spin orientation of the electrons in the $(1,1)$ state since the only accessible $(0,2)$ state is a spin singlet (Fig.\ \ref{fig:fig1}).

Magnetic spin resonance is achieved by sending an alternating current through a coplanar stripline (CPS) which lies on top of the surface gates, separated by a thin dielectric layer. This current produces a small oscillating magnetic field $B_1 \simeq 1$~mT perpendicular to the external magnetic field $B_0 \simeq 100$~mT. The experimental data are obtained with the same device and in the same measurement run as the data presented in Ref.~\cite{frank:nature}. The difference is that the device is tuned to a higher interdot tunnel coupling and coupling to the right lead.

When we apply a continuous wave RF current with fixed frequency $\omega$ to the CPS and sweep the external magnetic field $B_0$ passing the resonance condition $B_0 = B_\text{res} \equiv \hbar \omega/g\mu_B$, we make a remarkable observation. One would expect that the resonance manifests itself as a peak in the current~\cite{frank:nature}. Indeed, if the external field is swept from low to high values, the current jumps up upon achieving the resonance condition. Unexpectedly, this resonant response extends over a wide range of magnetic fields, that exceeds $B_\text{res}$ by a factor of $2$ (see Fig.~\ref{fig:exp}a upper panel). If the field is swept in opposite direction (Fig.~\ref{fig:exp}a lower panel), the current remains low till $B_0$ is several mT above $B_\text{res}$.. This indicates a strong hysteresis for $B_0 > B_\text{res}$, whereas the hysteresis below $B_\text{res}$ is much less pronounced.

Another unexpected observation is made at fixed $B_{0} \approx B_\text{res}$. Instead of a single value of the current corresponding to the maximum value of the ESR satellite peak, we observe clearly distinguishable {\em multiple} stable values of the current. Switching between these values gives rise to a random telegraph signal (RTS) with time scales ranging from seconds to minutes. Typical time-resolved measurements of the RTS are presented in Fig.~\ref{fig:exp}b for three different values of the energy level detuning $\Delta_{LR}$ (Fig.~\ref{fig:fig1}).

We associate both the hysteresis and RTS with DNSP induced by the non-equilibrium electron spin dynamics under conditions of ESR and transport in the dots. Nuclear polarization is known to provide an extra effective magnetic field $B_N$ acting on the electron spin~\cite{overhauser}. Where high current is observed in the hysteresis region, this extra field should be such that the total field $B_0+B_N\approx B_\text{res}$, i.e.\ the nuclear field `tunes' the system to the resonance condition~\cite{danon:056603}. Low current indicates that the total field $B_0+B_N$ significantly deviates from $B_\text{res}$: The nuclei are unpolarized. Both polarized and unpolarized states are stable in the interval of hysteresis. Fluctuations of any kind could provide spontaneous switching between stable states, leading to the RTS.

A number of experimental details does not fit into this simple picture. Firstly, there are {\em multiple} values of the current observed, three are clearly visible in Fig. \ref{fig:exp}b (labeled $A$-$C$). This implies multiple stable states of nuclear polarization with a total field close to $B_\text{res}$. Actually, we think that the RTS traces provide evidence for the existence of a fourth state. There is a number of current dips observed (labeled $D$) too big to be statistical fluctuations. We interpret those dips as signatures of a fourth state that decays on the scale of a second, i.e.\ different from state $A$, which decays on a larger time scale. Secondly, switching between the different current levels is rather fast. The nuclear spin dynamics are known to be slow, with a typical relaxation time $\tau_n \sim 15$~s~\cite{frank:science,maletinsky:056804,D.J.Reilly08082008}. If the current is a direct measure of the nuclear polarization, then why is the duration of the switching events so short? 

A third point is the {\em sign} of the polarization. Usually, in ESR experiments the dominating mechanism of DNSP is described by the Overhauser effect: The ESR excitation drives the electron spin(s) out of equilibrium, and hyperfine induced electron-nuclear spin exchange is one of the mechanisms contributing to electron spin relaxation. As reasoned by Overhauser, on grounds of spin conservation, the direction of nuclear polarization should be parallel to the spin of excited electrons, whatever its orientation is with respect to the magnetic field applied. This is the case for most DNSP experiments, e.g.~\cite{frank:nature,kroner:156803,tartakovskii:026806}. Given the negative $g$-factor and positive hyperfine coupling in GaAs~\cite{paget:1977}, this would give a $B_N$ {\em parallel} to $B_0$~\cite{danon:056603}. In our experiment, its direction is clearly {\em opposite}, as high current is seen for $B_0 > B_\text{res}$. All three points are captured by the theory given below.

%\section{Ingredients}

The electron spin $\mathbf{\hat S}$ and nuclear spins $\mathbf{\hat I}_k$ in each dot are coupled by hyperfine interaction~\cite{paget:1977}
\begin{equation}\label{eq:hfham}
\hat H_{\text{hf}} = \frac{1}{2} \sum_k A_k\left\{ 2\hat S^z \hat I^z_k + \hat S^+ \hat I^-_k + \hat S^-\hat I^+_k \right\},
\end{equation}
where the sum runs over all $N\sim 10^6$ nuclei in the dot. The energy $A_k$ is proportional to the probability to find the electron at the position of nucleus $k$,  $A_k \simeq 10^{-10}$~eV. With an external field applied in the $z$-direction, the `flip-flop' terms $\hat S^\pm \hat I_k^\mp$ provide spin exchange between the electrons and nuclei. Owing to energy conservation, these exchange transitions must be second-order processes involving a mechanism supplying or absorbing the excess Zeeman energy. Conventionally, the electron-nuclear spin exchange is due to the time-independent hyperfine coupling $A_k$. However, as recently has been pointed out~\cite{frank:nature,laird:246601}, in this setup a significant a.c.\ electric field moves the electrons in the dots with respect to the nuclei. This can be accounted for by introducing a {\em time-dependent} component in the hyperfine coupling $ A_k \to A_k + \tilde A_k e^{i\omega t} + \tilde A^{*}_k e^{-i\omega t}$. We estimate that under the present conditions $\tilde A_k /A_k \simeq 0.1$~\cite{laird:246601,Rudner}.

We have considered six candidate mechanisms for DNSP~\cite{supp}, assuming a saturated ESR. We concluded that the dominant one involves the time-dependent hyperfine coupling, which allows for 'photon assisted flip-flops'. These flip-flops not have a preferred direction set by a large energy mismatch: now the spin asymmetry is now provided by internal spin relaxation causing the spin ground state (parallel to the external field) to be more populated than the excited state.

%\section{Model}

The theoretical consideration includes the following steps:
(i) We consider the four $(1,1)$ states using a rotating wave approximation, assuming a saturated ESR and a negligible exchange splitting, i.e.\ $\text{min}\{t,t^2/\Delta_\text{LR}\} \ll B_1, B_N$. The eigenstates in a rotating frame are mixtures of spin-up and spin-down states, with a mixing angle $\theta_{L,R} =\frac{1}{2} \arctan\{\tilde B_{L,R} / 2 f_{L,R}\}$ which can be different in both dots (see Fig.~\ref{fig:fig1}), due to e.g.\ different coupling of the electrons to the CPS. The Rabi frequency in each dot $\tilde B_{L,R} \equiv g\mu_B B^{(L,R)}_1/\hbar$ gives the width of the saturated resonance, and the ESR frequency mismatch $f_{L,R} \equiv |g\mu_B(B_0+B_N^{L,R})/\hbar|-\omega$ depends on the nuclear polarization in each dot.
(ii) We evaluate the transition rates between these states to obtain their quasi-stationary population and the current through the double dot. We include tunneling (characterized by $\Gamma_s =t^2/\Gamma_\text{out} \simeq 1-10~$MHz) and single electron spin relaxation~\cite{danon:056603} ($\propto \Gamma_r \simeq 1~$MHz at zero temperature, which will be enhanced by a thermal factor $k_B T /g\mu_BB_0\equiv \xi \simeq 5$, in accordance with a lower bound estimate set by the typical leakage current of 100 fA). This approach is valid in the limit $\tilde{B} \gg \Gamma_{s,r}$.
(iii) We compute the rates of hyperfine-induced spin exchange. In the first approximation we find rates symmetric with respect to nuclear spin, their scale set by $\Gamma_2 \simeq \tilde A_k^2/(64\hbar^2\xi\Gamma_r) \sim 0.5$~Hz. Being symmetric, these rates do not contribute to DNSP. They merely enhance the relaxation of the nuclear fields.
(iv) The small spin-asymmetric part of these rates $\Gamma_1 \simeq \frac{5}{3}(\tilde A_k/8\hbar\tilde{B})^2 (\Gamma_s/\xi) \sim 10^{-2}$~Hz, due to electron spin relaxation, introduces a preferential direction of nuclear spin pumping in each dot.
(v) We construct equations of motion for the effective nuclear fields $B_N^{L,R}$ and analyze the stable states of nuclear polarization given by $d{B}_N^{L,R}/dt=0$.
(vi) We use a Fokker-Planck equation to give a qualitative analysis of fluctuations of nuclear polarization and switching rates between the stable states.

\begin{figure}[b]
\includegraphics[width=8.5cm]{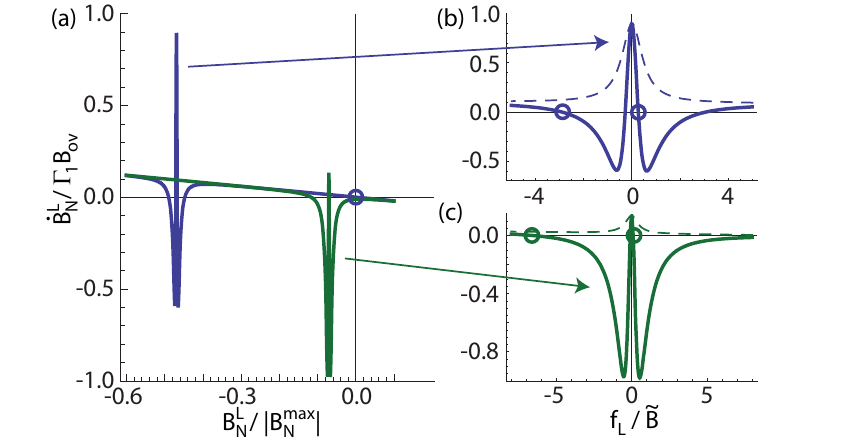}
\caption{(a) Time-derivative $d B_N^L/dt$ at the edge of the hysteresis interval $B_0 \approx B_\text{res}$ (green) and in the middle of the interval $B_0 \approx B_\text{res}+ 0.5\ |B_N^\text{max}|$ (blue). (b,c) Close-up at resonance. The curves consist of the usual relaxation (linear slope) which is resonantly enhanced (dashed lines), and spin pumping that adds a two-peak shape near the resonance. The circles indicate the stable states of nuclear polarization. We used $\Gamma_1 / \Gamma_2 = 0.043$, $\Gamma_2\tau_n = 5$, $\theta_R =0$, $\xi\Gamma_r/\Gamma_s = 0.75$, and assumed equally strong coupling $\tilde A_k$ of all nuclei to the electron.} \label{fig:plot}
\end{figure}

\begin{figure*}
\includegraphics[width=17cm]{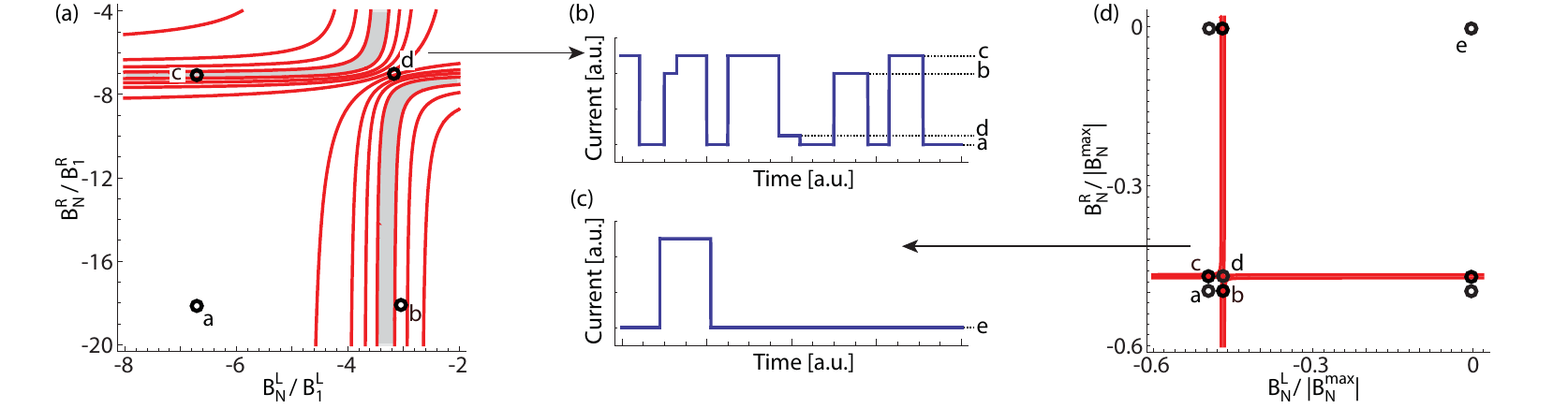}
\caption{Stable polarizations in the plane ($B_N^L,B_N^R$), for the cases (a) $B_0-B_\text{res}\sim B_1$ and (d) $B_0-B_\text{res}\sim 0.5\ |B_N^\text{max}|$. A contour plot of the current is included, the gray shade indicating the region with highest current. Switching between the stable points gives rise to RTS as presented in (b) and (c). A qualitative difference is that the point $e$ in (d) is `isolated', i.e.\ having switched to $e$, the system will never switch back. In (a) an asymmetry in $\tilde B_{L,R}$ and $N_{L,R}$ is included, resulting in four different current levels for $a$-$d$, whereas (d) is plotted assuming a symmetric double dot. Note the different scales at the axes in (a) and (d). The same plots (a) and (d) can be found in the Supplementary Material where we included the local nuclear spin dynamics as a vector field.} \label{fig:fig4}
\end{figure*}

The evolution equation for $B_N^L$ thus found reads
\begin{equation}\label{Kdot}
\frac{dB_N^L}{dt} = -\Gamma_1 B_\text{ov} P(\theta_{L,R})-\left\{ \frac{1}{\tau_n} + \Gamma_2 R(\theta_{L,R})\right\} B_N^L,
\end{equation}
and the equation for $B_N^R$ is obtained by permutation of $L$ and $R$. The field $B_\text{ov}$ is the Overhauser field of full polarization, $B_\text{ov} \approx 5$~T for GaAs. The functions $P$ and $R$ are dimensionless functions giving the functional dependence of the resonant nuclear spin pumping ($P$) and resonantly enhanced nuclear spin relaxation ($R$) on the mixing angles and on $\Gamma_s/\xi\Gamma_r$, and have a maximum $\sim 1$. While $R$ is roughly Lorentzian-shaped, the function $P$ is zero far from resonance $\theta \to \{ 0,\pi \}$, reaches maximum at $|f| \simeq \tilde{B}$, and falls off to zero again at the resonance $\theta = \pi/2$. This resonant dip is due to the vanishing of electron spin polarization at the saturated resonance. In Eq.\ (\ref{Kdot}), the terms proportional to $-B_N$ give nuclear spin relaxation: The first term presents the usual $\tau_n$ while the second term gives a resonant enhancement owing to spin exchange with electrons. Nuclear spin pumping is given by $\Gamma_1 B_\text{ov} P$ ($\sim 50$~mT/s, much faster than the sweep rate in Fig.\ \ref{fig:exp}a), with a \emph{sign opposite} to that following from the Overhauser reasoning: Spin exchange under conditions of electron transport is mostly due to electrons polarized along the direction of the external field. The shape of a typical pumping curve is shown in Fig.\ \ref{fig:plot}.

We are now also able to understand the extended interval of hysteresis: ESR response can be observed as long as there exist stable solutions of $dB_N/dt=0$ close to resonance. Eq.\ (\ref{Kdot}) determines the interval of hysteresis as $B_\text{res} \lesssim B_0 <B_\text{res} +|B_N^\text{max}|$, where the maximal nuclear field is $B_N^\text{max} = -B_\text{ov}\Gamma_1/(\Gamma_2+\tau_n^{-1})$. Using the parameters as estimated above we find that $\Gamma_2 \tau_n \sim 10$.

It is the two-peak shape of the pumping curve that is responsible for the multiple stable states of nuclear polarization, even at the edge of the hysteresis interval. If $B_0\approx B_\text{res}$ (Fig.\ \ref{fig:plot}, green curve), there are four stable states for the double dot system. This is represented in Fig.\ \ref{fig:fig4}a, where the circles indicate the stable points in the plane $(B_N^L,B_N^R)$. It is now clear how, even close to $B_0=B_\text{res}$, the system can have \emph{four stable states} with different current. A rough estimate for the duration of the switching between those states is the typical distance ($\sim B_1$) over the local speed of the spin dynamics ($\sim \Gamma_1 B_\text{ov}$), giving $\sim 10^{-2}$~s, which explains the \emph{fast switching}. A typical time trace in this case will look like Fig.\ \ref{fig:fig4}b, which is to be compared with Fig.\ \ref{fig:exp}b.

When increasing $B_0$, both dots will develop a separate third unpolarized stable state (Fig.\ \ref{fig:plot}, blue curve), giving as many as nine stable points, as presented in Fig.\ \ref{fig:fig4}d. At higher fields the unpolarized state (labeled $e$) will become isolated from the other stable states: If the system switches to $e$, it will never switch back (see Fig.\ \ref{fig:fig4}c). This also has been observed in experiment~\cite{ivo:pc}. When subsequently sweeping back from high to low field, the barrier for switching back from $e$ to a high-current state is again gradually lowered. When the typical switching time becomes comparable to the time scale of the sweep, one can expect the current to switch to a high value (Fig.\ \ref{fig:exp}a, lower panel).

From Eq.\ (\ref{Kdot}) we construct a two dimensional Fokker-Planck equation to study the stochastic properties of the polarizations in more detail~\cite{danon:056603}. Importantly, due to the accelerated dynamics, the fluctuations around all polarized states are suppressed as $\langle (\Delta B_N)^2\rangle/\Omega^2 \approx (B_1/|B_N^\text{max}|)$, $\Omega^2 \equiv (A_k/g\mu_B)^2N$ being the field variance in the unpolarized state.
%did not define N yet
Using Kramers' method~\cite{vankampen} we derive an expression for the switching rates between the stable states. All rates have the exponential dependence $\Gamma_\text{sw} \propto \exp \{-\alpha B_1 |B_N^\text{max}|/\Omega^2 \}$, where $\alpha$ is a numerical factor: The rates are suppressed exponentially with a power $\sim B_1^2 / \langle (\Delta B_N)^2\rangle \gg 1$. This exponential dependence explains the large RTS time scale as well as the strong variation with $\Delta_{LR}$ in Fig.\ \ref{fig:exp}. We calculated the exponent explicitly for $\Gamma_\text{sw}$ from $a$ to $d$ in Fig.\ \ref{fig:fig4}d. We used $\Gamma_2\tau_n = 10$, $\Gamma_s/\xi\Gamma_r = \frac{16}{15}$, and $B_0-B_\text{res} = 0.5\ |B_N^\text{max}|$ and found that $\alpha \approx 0.72$.

To conclude, we have observed multiple nuclear polarization states and locking of the ESR condition over a large range of magnetic fields in a double quantum dot under ESR. We presented a theoretical model that captures the existence of these phenomena and their unusual features as fast switching and a `wrong' sign of DNSP. We acknowledge useful discussions with M.\ Laforest. This work was supported by the Dutch Foundation for Fundamental Research on Matter (FOM).

\onecolumngrid
\newpage

\setcounter{page}{1}
\thispagestyle{empty}

\begin{center}
\textbf{{\large Supplementary Material for\\ ``Multiple Nuclear Polarization States in a Double Quantum Dot''}}\\
\bigskip
J.\ Danon, I.T.\ Vink, F.H.L.\ Koppens, K.C.\ Nowack, L.M.K.\ Vandersypen, and Yu.V.\ Nazarov\\
\textit{Kavli Institute of NanoScience, Delft University of Technology, 2628 CJ Delft, The Netherlands}
\end{center}

\section{Sample}

\begin{figure}[htp]
\includegraphics[width=10cm]{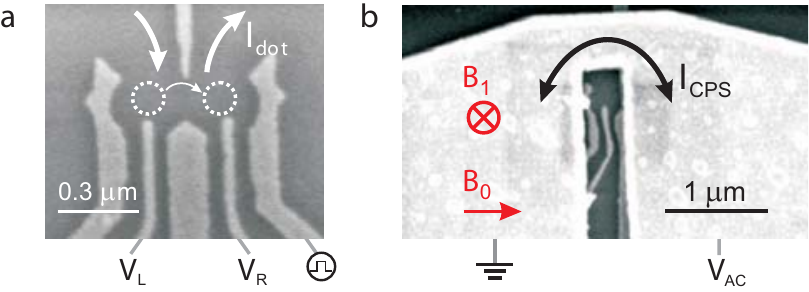}
\caption{ESR device. (a) Scanning electron microscope (SEM) image of a device with the same gate pattern as used in the experiment. The Ti/Au gates are deposited on top of a GaAs/AlGaAs heterostructure containing a two-dimensional electron gas 90 nm below the surface. White arrows indicate current flow through the two coupled dots (dotted circles). The directions of the external magnetic field and the ac magnetic field are indicated. (b) SEM image of a device similar to the one used in the experiment. The termination of the coplanar stripline is visible on top of the gates. The gold stripline has a thickness of 400 nm and is designed to have a 50~$\Omega$ characteristic impedance, $Z_\mathrm{0}$, up to the shorted termination. It is separated from the gate electrodes by a 100-nm-thick dielectric (Calixerene).}
\label{esrdevice}
\end{figure}

The experimental data presented are obtained with the same sample as used in reference \cite{koppens06}. A device with the same gate pattern as used in the experiment is shown in Fig.\ \ref{esrdevice}a. The two coupled semiconductor quantum dots are defined by surface gates (Fig.\ \ref{esrdevice}a) on top of a two-dimensional electron gas (2DEG). The oscillating magnetic field that drives the spin transitions is generated by applying a radio-frequency (RF) signal generated by a Rohde \& Schwarz SMR40 source to an on-chip coplanar stripline (CPS) which is terminated in a narrow wire, positioned near the dots and separated from the surface gates by a 100-nm-thick dielectric (Fig.\ \ref{esrdevice}b). The current through the wire generates an oscillating magnetic field $B_1$ at the dots, perpendicular to the static external field $B_0$ and slightly stronger in the left dot than in the right dot.

The GaAs/AlGaAs heterostructure from which the samples were made was purchased from Sumitomo Electric. The 2DEG has a mobility of $185 \times 10^3$~cm$^2/$Vs at 77~K, and an electron density of $4$-$5 \times 10^{11}$~cm$^{-2}$, measured at 30~mK with a different device than used in the experiment.

Background charge fluctuations made the quantum dot behavior excessively irregular. The charge stability of the dot was improved considerably in two ways. First, the gates were biased by +0.5~V relative to the 2DEG during the device cool-down. Next, after the device had reached base temperature, the reference of the voltage sources and I/V converter (connected to the gates and the 2DEG) were biased by +2~V. This is equivalent to a $-2$~V bias of both branches of the CPS, which therefore (like a gate) reduces the 2DEG density under the CPS.

The measurements were performed in a Oxford Instruments Kelvinox 400 HA dilution refrigerator operating at a base temperature of 35-40~mK.

\section{Measurements}

In both traces in Fig.\ 2a in the main text, the nuclear bath is unpolarized at the onset of electron spin resonance. In the case of the upper panel, $B_0$ is swept just before the measurement from 300~mT to $-20$~mT in about 40~s. During this sweep the nuclear field relaxes (typical relaxation time $\tau_n \sim$ 10~s) or is even actively depolarized. Residual polarization would be indicated by a shift of the zero field peak and the onset of ESR response to a nominal non-resonant magnetic field, which is both not observed. In the case of the lower panel the magnetic field is swept from low to high magnetic field (to 300~mT) just before recording the trace. In this case there could be polarization still present at the beginning of the trace, however in that case that polarization relaxes much faster than the sweep rate of 60~mT/min, such that when reaching the resonant field the nuclear spin bath is equilibrated.

\section{Candidate Mechanisms}

\begin{figure}[b]
\includegraphics[width=17cm]{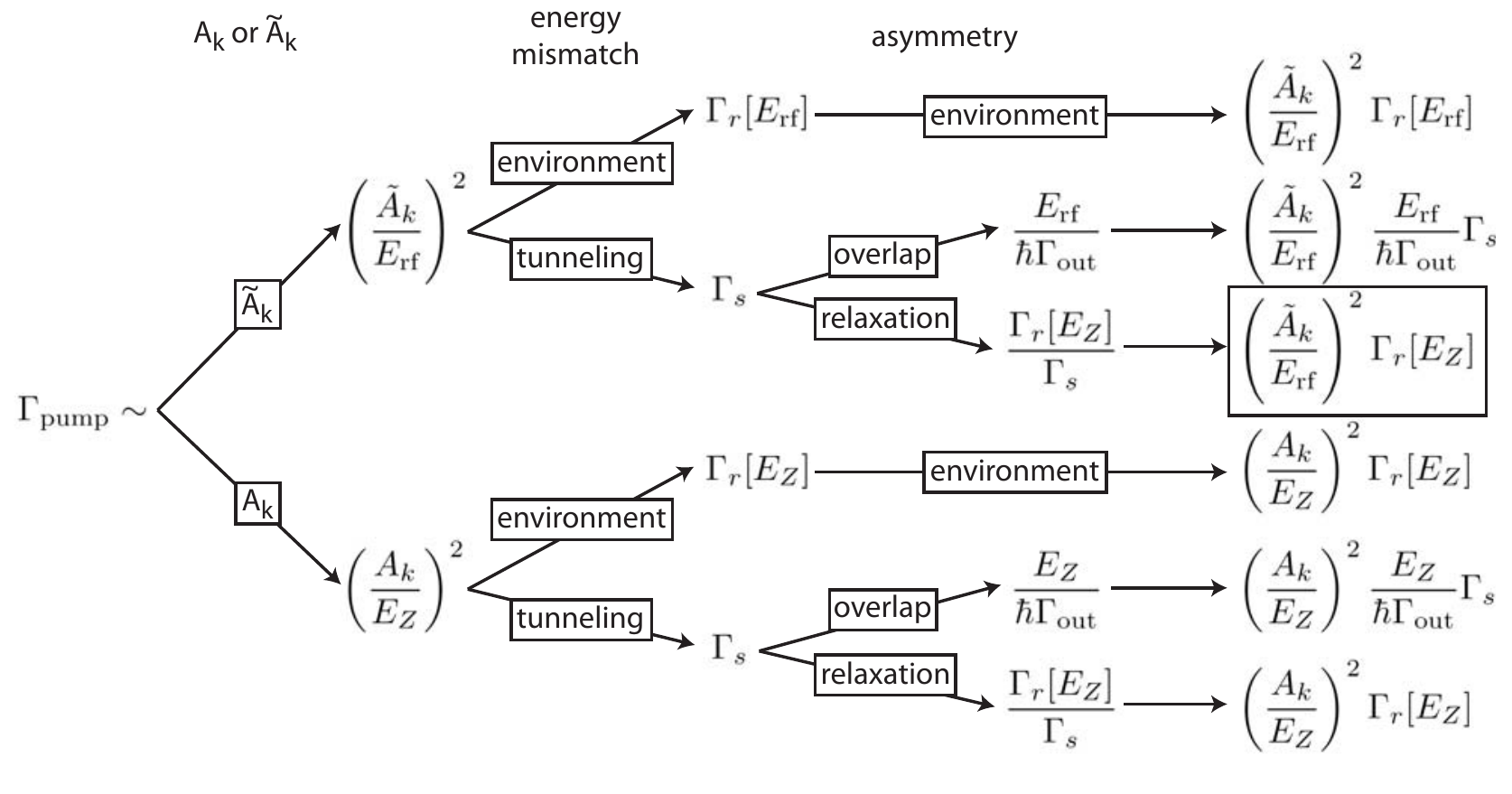}
\caption{Overview of all mechanisms considered and the corresponding estimates for the nuclear spin pumping rate. The two key ingredients for spin pumping are dissipation of the energy mismatch and an asymmetry in spin giving a preferred direction for nuclear spin flips. Furthermore, we considered both the effect of $\tilde A_k$ and $A_k$, i.e.\ time-dependent and time-independent hyperfine coupling. We conclude that, under the present experimental conditions, the dominant mechanism involves time-dependent hyperfine coupling, energy dissipation by electron transport and internal spin relaxation causing an asymmetry in the populations of the electron spin states.} \label{fig:mech}
\end{figure}

Here we describe how we identify the dominating process of hyperfine induced nuclear spin flips. The `flip-flop' terms $\hat S^\pm \hat I_k^\mp$ in the hyperfine Hamiltonian are responsible for the exchange of spin between the electron and the nuclei. However, as the nuclear Zeeman splitting is 3 or 4 orders smaller than the electron splitting~\cite{hanson:1217}, the states coupled by $\hat S^\pm \hat I_k^\mp$ are roughly $g\mu_BB_0$ apart in energy. Therefore, spin exchange is only allowed in a second-order process in which some other mechanism supplies or absorbs the excess Zeeman energy.

This energy difference may (i) be dissipated by an environment~\cite{danon:056603s}, or (ii) be given to an electron tunneling out of the dot. In case (i) the environment, at sufficiently low temperatures, can only absorb energy, so that the electron Zeeman energy can only be reduced. This results in the same sign of DNSP as with the usual Overhauser effect~\cite{overhausers}. In case (ii), owing to a voltage bias much larger than the Zeeman energy, the change of energy
in the course of a spin-flip can be of either sign. In this case, a preferential direction of DNSP will be determined by some other spin asymmetry of the system. Such an asymmetry may arise from either (ii.a) a difference in spin-flip rates for different spin directions (e.g.\ due to different overlap between initial and final states), or (ii.b) different populations of the states with different spin directions (e.g.\ due to internal relaxation processes or differing decay rates).

Apart from these three mechanisms, there are two more options to choose between: As mentioned in the main text, the a.c.\ electrical component of the exciting field $B_1$ moves the electrons in the dots with respect to the nuclei, and this we account for by introducing a time-dependent component in the hyperfine coupling. The time-dependent and time-independent couplings will give rise to different flip rates, so this gives us in total six candidate mechanisms.

Let us first decide on the relative contributions of the time-dependent and time-independent hyperfine couplings, $A_k$ and $\tilde A_k$. We compare the strength of second order transition rates, in both cases proportional to the
coupling amplitude square and inversely proportional to the energy square of the virtual state. While for the time-independent coupling this energy is the electron Zeeman energy $E_Z \equiv g\mu_BB_0$, it is a much smaller energy
for the resonant time-dependent coupling, involving the a.c.\ resonant magnetic field $E_\text{rf} \equiv g\mu_BB_1$. Therefore we have to compare the factors $(A_k/E_Z)^2$ and $(\tilde A_k/E_\text{rf})^2$. We estimate that for our conditions $\tilde A_k /A_k \simeq 0.1$ and $E_\text{rf}/E_Z = B_1/B_0 \simeq 0.01$, and conclude that the time-dependent coupling dominates.

To decide upon the other three options mentioned above, we have to compare the spin exchange rates involving electron tunneling, characterized by the broadening of the $(0,2)$ singlet $\hbar\Gamma_\text{out}$ and the typical decay rate of the $(1,1)$ singlet $\Gamma_s\simeq t^2/\Gamma_\text{out}$, and internal spin relaxation within the dots, characterized by a rate $\Gamma_r[\varepsilon]$, $\varepsilon$ being the energy dissipated. For mechanism (i) we find the scale $(\tilde A_k/E_\text{rf})^2\Gamma_r[E_\text{rf}]$, i.e.\ spin relaxation dissipates the remaining energy difference $\sim E_\text{rf}$. In case (ii) the energy is dissipated during tunneling, which takes place with a rate $\sim \Gamma_s$, giving a scale for the nuclear spin flip rate of $(\tilde A_k/E_\text{rf})^2\Gamma_s$. This rate however is symmetric in spin direction, so to find a preferred direction of DNSP we need to include an asymmetry: (ii.a) The states are split by $\sim E_\text{rf}$, so decay to the broadened $(0,2)$ singlet introduces a relative difference of $\sim E_\text{rf} / \hbar\Gamma_\text{out}$ in the rates, setting the scale of the DNSP rate $\sim \tilde A^2_k \Gamma_s/ E_\text{rf} \hbar \Gamma_\text{out}$. (ii.b) Internal spin relaxation competes with tunneling processes, causing an asymmetry in the population probabilities of the states of $\sim \Gamma_r [E_Z]/\Gamma_s$ resulting in $(\tilde A_k/E_\text{rf})^2 \Gamma_r [E_Z]$ for DNSP.
In Fig.\ \ref{fig:mech} we give a schematic representation of these considerations. We show all mechanisms investigated and give the corresponding estimates of the scale of nuclear spin pumping.

We estimate $\tilde A_k \sim 10^{-11}$~eV, $E_\text{rf} \sim 10^{-8}$~eV, $\Gamma_s \sim 10$~MHz, $\hbar \Gamma_\text{out} \sim 10^{-4}$~eV and $\Gamma_r[E_Z]\sim 100\times\Gamma_r[E_\text{rf}]\sim 1$~MHz, resulting in the estimates for the scales of DNSP rate (i) $10^{-2}$~Hz, (ii.a) $10^{-4}$~Hz, and (ii.b) $1$~Hz. Based on this argument we conclude that mechanism (ii.b) dominates: Electric field assisted hyperfine flip-flops involve the absorption and emission of photons with energy $\hbar\omega$. Close to resonance this effectively reduces the energy mismatch of the states involved in a flip-flop from $g\mu_BB_0$ to the energy scale of the ESR driving $g\mu_BB_1$. Since this energy mismatch is too small to result in a significant nuclear spin pumping rate based on a standard Overhauser argument, another spin asymmetry is needed. Internal electron spin relaxation provides this asymmetry: it causes the electron spin ground state to be (slightly) more populated than the excited state. This difference in populations combined with photon assisted hyperfine flip-flops (which do not have a preferred direction) results in DNSP parallel to the spin of the electron ground state.

\section{Theory}

Here we will elaborate further on the six steps of the theoretical consideration as sketched in the main text.

(i) The Hamiltonian for the electron spin operators $\mathbf{\hat S}_{L,R}$ in the rotating wave approximation reads
\begin{equation}\label{eq:ham0}
\hat H = -\hbar f_{L} \hat S^z_{L}-\hbar f_{R} \hat S^z_{R} + \frac{\hbar}{2} \left( \tilde B_{L} \hat S^x_{L}+\tilde B_{R} \hat S^x_{R}\right),
\end{equation}
$L(R)$ referring to the left(right) dot. The rotating wave approximation is justified by $\tilde{B},|f|\ll \omega$. The eigenstates of $\hat H$ form the basis $\{ \ket{+}_L,\ket{-}_L \} \otimes \{ \ket{+}_R,\ket{-}_R \}$, with $\ket{+} = \cos \theta \ket{\ua} + \sin \theta \ket{\da}$ and $\ket{-} = \sin \theta \ket{\ua} -\cos \theta \ket{\da}$, where the mixing angle is $\theta_{L,R} =\frac{1}{2} \arctan\{\tilde B_{L,R} / 2 f_{L,R}\}$.

(ii) The master equation includes the decay and relaxation rates, and is justified if $\tilde{B}$ by far exceeds these rates~\cite{Rudners}. The rates depend on the wave functions of the states involved. Any basis state $\ket{n} \in \{\ket{++},\ket{+-},\ket{-+},\ket{--} \}$ decays via the $(0,2)$ singlet to $(0,1)$ with a rate $\Gamma_s^n = \left| \braket{S | n}\right|^2\Gamma_s$, with $\ket{S}$ being $(1,1)$ singlet. Such a decay process is followed by a charge transfer in the left junction $(0,1) \to \ket{m}$, whereby all four basis states $\ket{m}$ are re-initialized with equal rates $\Gamma_s/4$. Internal relaxation processes are due to coupling to an environment and involve energy dissipation of $\pm E_Z$~\cite{danon:056603s}. We believe that the environment are mainly the electrons in the leads. Their temperature is typically large, $\xi \equiv k_BT/E_Z \simeq 5 \gg 1$, so we need to consider both emission and absorption rates. They read $\Gamma_{\text{abs}} = n_B(E_Z)\Gamma_r[E_Z]$ and $\Gamma_{\text{em}} = \Gamma_\text{abs} + \Gamma_r[E_Z]$, with $n_B(\varepsilon)$ being the Bose distribution and $\Gamma_r[E_Z]$ being the emission rate at zero temperature. In the high-temperature limit we find the transition rates $\Gamma_r^{n\to m} \approx \{ \xi - \sum_{L,R}| \bra{m} \hat S_{L,R}^-\ket{n} |^2 \}\Gamma_r[E_Z]$. We are now able to construct a master equation
\begin{equation}
0 = -\Gamma_s^n p_n + \frac{1}{4}\sum_m \Gamma_s^m p_m + \sum_m \left\{\Gamma_r^{m\to n}p_m - \Gamma_r^{n\to m} p_n \right\},
\label{eq:master}
\end{equation}
and solve it for the quasi-stationary populations $p_n$. These populations gain, via the rates $\Gamma_s^n$ and $\Gamma_r^{n\to m}$, a resonant dependence on $f_{L,R}$ on the scale $f \simeq \tilde B$ and therefore also depend on the nuclear polarizations $B_N^{L,R}$. From the populations $p_n$ we can calculate the current through the double dot as $I_\text{dot} = e\Gamma_s \sum_{n} |\langle S | n \rangle |^2 p_n$.

(iii-v) The rates of electron-nuclear spin exchange are calculated using second order perturbation theory. The positive and negative spin flip rates \emph{per nucleus} in the left(right) dot read~\cite{endnote1}
\begin{equation}\label{eq:2ndorder}
\Gamma^{(1)}_{\pm,L(R)}= \frac{1}{16} \tilde A_k^2 \Gamma_s \sum_{n,m} \left|\frac{\braket{S\,|\,m}\bra{m}\hat S_{L(R)}^\mp\ket{n}}{E_n - E_m} \right|^2 p_n.
\end{equation}
Non-zero diagonal matrix elements such as $\bra{++} \hat S^\pm \ket{++}$, will give rise to very small denominators in (\ref{eq:2ndorder}), of the order of the nuclear Zeeman energy. Therefore, we have to investigate the contribution of these, possibly dominating, terms in another way. We write the second order perturbation in the hyperfine Hamiltonian,
\begin{equation}
\frac{d\rho}{dt} = - \int^t \left[ \hat H_L'(t)+\hat H_R'(t),\left[ \hat H_L'(t')+\hat H_R'(t'), \rho\right] \right] dt',
\end{equation}
where the perturbation is $\hat H_{L(R)}'(t) = \frac{1}{4} \sum_k \tilde A_k \{\hat S_{L(R)}^+(t) \hat I^-_{k,L(R)}(t) + \hat S_{L(R)}^-(t)\hat I^+_{k,L(R)} (t) \}$. After separating the time scales of the electronic and nuclear spin dynamics, assuming that we can separate the electronic and nuclear part of the density matrix as $\rho = \rho_\text{el}\otimes\rho_\text{nuc}$, and tracing over the electron part of the density matrix, we find that we can write for the time-evolution of the nuclear field in one of the dots
\begin{equation}\label{eq:corr}
\frac{dB_N}{dt} = \frac{\tilde A_k^2}{8\hbar^2} \left\{ \frac{5}{3}(\chi^{xy}-\chi^{yx}) B_\text{ov} - (R^{xx}+R^{yy}) B_N \right\},
\end{equation}
where $\chi^{ab} = -i\int^t \langle \hat S^a(t)\hat S^b(t') - \hat S^b(t')\hat S^a(t)\rangle dt'$ and $R^{ab} = \int^t \langle \hat S^a(t)\hat S^b(t')+\hat S^b(t')\hat S^a(t) \rangle dt'$, i.e.\ the susceptibility and zero-frequency fluctuations of the electron spin in the dot under consideration. In Eq.\ (\ref{eq:corr}) we left out the contributions proportional to the polarization in the $x$- and $y$-direction while they are averaged out to zero. We focus on the contributions of the diagonal matrix elements and find that $\chi^{xy}-\chi^{yx} = 0$ and $R^{xx}+R^{yy}$ is only non-zero close to resonance, resulting in a resonant enhancement of nuclear spin relaxation.

\begin{figure}[t]
\includegraphics[width=17cm]{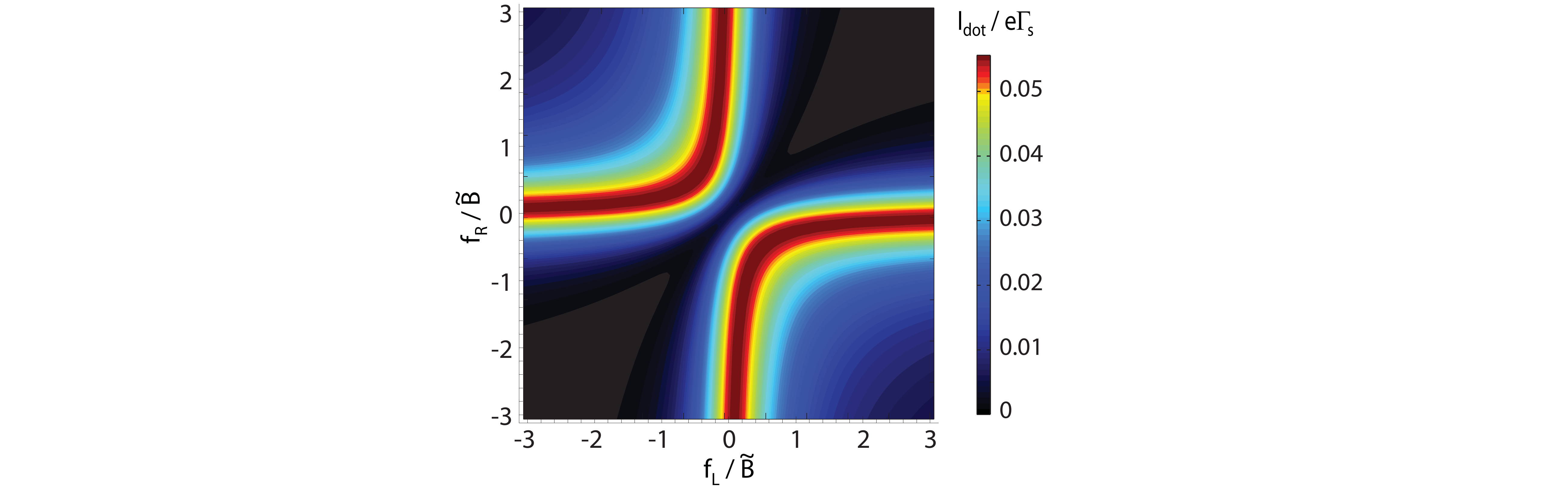}
\caption{The current through the double quantum dot $I_\text{dot}$ as a function of $f_L$ and $f_R$. We find high current when only one of the two dots is on resonance and low current in the rest of the plane. To generate this plot we used $\xi=5$ and $\Gamma_s = 20 \Gamma_r[E_Z]$ and we subtracted the leakage current far away from both resonances.} \label{fig:fig1s}
\end{figure}
We can combine Eqs (\ref{eq:2ndorder}) and (\ref{eq:corr}) in an evolution equation for the nuclear fields
\begin{equation}
\frac{dB_N^L}{dt} = -\Gamma_1 B_\text{ov} P(\theta_{L,R})-\left\{ \frac{1}{\tau_n} + \Gamma_2 R(\theta_{L,R})\right\} B_N^L,
\label{eq:kdot}
\end{equation}
where we added a term describing diffusive spin relaxation $\propto 1/\tau_n$. The equation for $B_N^R$ is obtained by permutation of $L$ and $R$. The scales $\Gamma_{1,2}$ are the same as defined in the main text, i.e.\ $\Gamma_1 = \frac{5}{3}(\tilde A_k/8\hbar\tilde{B})^2 (\Gamma_s/\xi)$ and $\Gamma_2 \simeq \tilde A_k^2/(64\hbar^2\xi\Gamma_r)$, and the functions $P$ and $R$ read
\begin{eqnarray}
P(\theta_{L,R}) & = & \frac{32\alpha(1+4\alpha)\cos \theta_L\cos\theta_R\cos (\theta_L-\theta_R)\sin^2 \theta_L + 4\alpha\{ 1+8\alpha +\cos^2(\theta_L-\theta_R)\}\sin^2 2\theta_L}{16\alpha(1+4\alpha)+\sin^2(\theta_L-\theta_R)} \\
R(\theta_{L,R}) & = & \frac{16 \alpha \sin^2 \theta_L}{1+16\alpha}\cdot
\frac{\cos^2(\theta_L-\theta_R)+1+8\alpha (3+16\alpha)}{\sin^2(\theta_L-\theta_R)+16\alpha (1+4\alpha)},
\label{eq:pr}
\end{eqnarray}
where $\alpha$ is the dimensionless variable $\alpha \equiv \xi\Gamma_r[E_Z]/\Gamma_s$. For this representation of $P$ we used the high temperature limit, i.e.\ $\xi \gg 1$ and assumed for simplicity all electron-nuclear spin couplings $\tilde A_k$ equal.

(vi) We also investigated both the switching rates between the different stable states, and the small fluctuations near these states. To estimate the fluctuations, we use a two-dimensional Fokker-Planck equation for the distribution function of the nuclear fields ${\cal P}(B_N^L,B_N^R)$, where $-1\leq B_N^{L,R}/B_\text{ov} \leq 1$. To derive the equation, we regard the nuclear dynamics in both dots as a random walk on a discrete set of spin values $n=\frac{1}{2} (N_\uparrow-N_\downarrow)$, where $N_{\ua(\da)}$ is the number of nuclei with spin up(down). The DNSP rate $\Gamma_1$ only causes transitions from $n$ to $n+1$, while the spin relaxation rates $\Gamma_2$ and $1/\tau_n$ cause transitions in both directions with a rate $(1/2\tau_n+\Gamma_2/2) N_{\uparrow,\downarrow}\gg \Gamma_1 N$, with $N\equiv N_\ua+N_\da$. We go to the continuous limit, justified by the large number of nuclei per dot ($N\sim 10^6$) to obtain~\cite{vankampens}
\begin{equation}\label{eq:fp}\begin{split}
\frac{\partial \mathcal{P}(B_N^L,B_N^R,t)}{\partial t} = &
\frac{\partial}{\partial B_N^L} 
\left\{- \mathcal{P} \frac{dB_N^L}{dt} + 
\frac{2B_\text{ov}^2}{N}\frac{\partial}{\partial B_N^L}\mathcal{P}\left(\frac{1}{2\tau_n} + \Gamma_2 \right) \right\} \\
& +\frac{\partial}{\partial B_N^R} 
\left\{- \mathcal{P} \frac{dB_N^R}{dt} + 
\frac{2B_\text{ov}^2}{N}\frac{\partial}{\partial B_N^R}\mathcal{P}\left(\frac{1}{2\tau_n} + \Gamma_2 \right) \right\}
\end{split}\end{equation}
\begin{figure}[t]
\includegraphics[width=17cm]{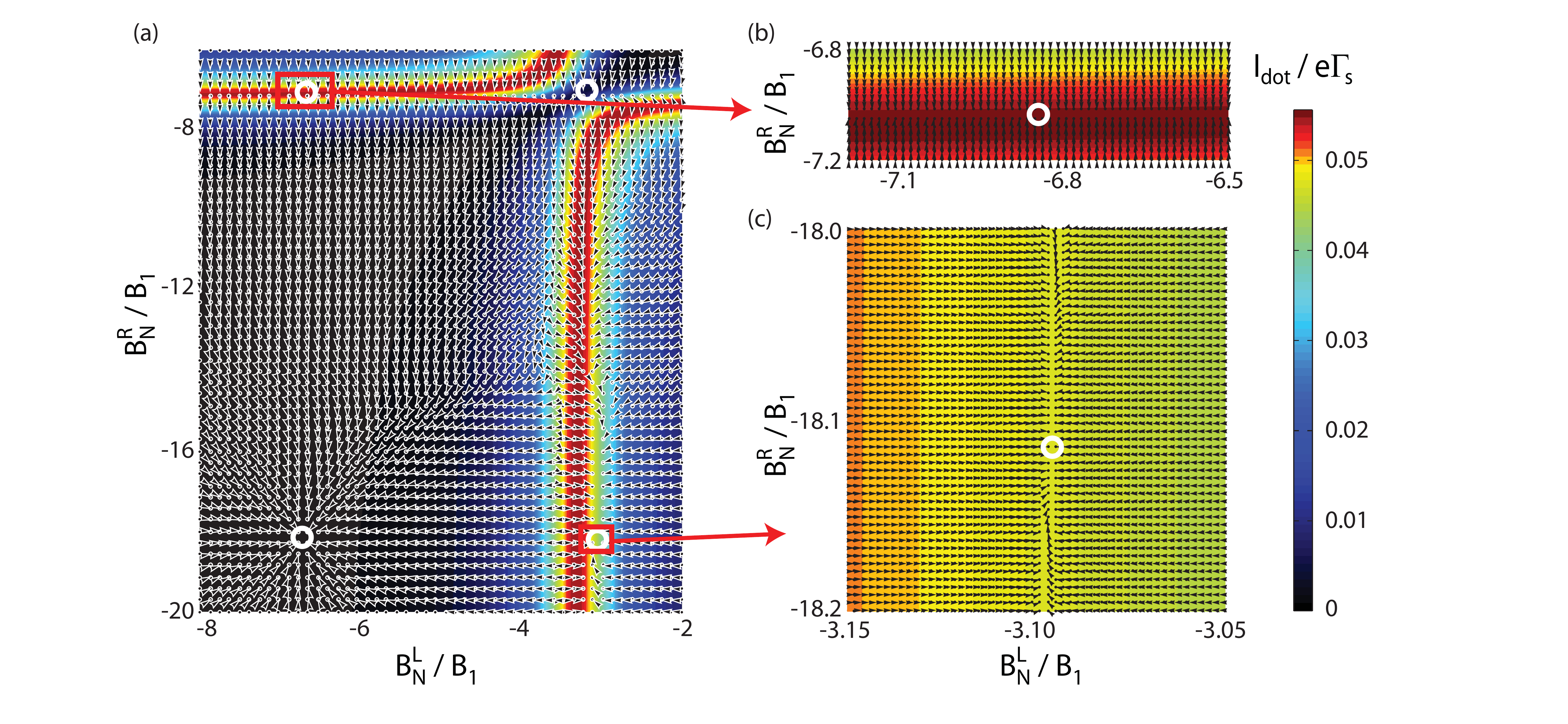}
\caption{On the edge of the hysteresis interval, $B_0-B_\text{res}\sim B_1$, assuming an asymmetric double dot (Fig.\ 4a in the main text). Here we add vector plots of $\{ \dot B_N^L, \dot B_N^R \}$ in the plane $(B_N^L, B_N^R)$ and the current $I_\text{dot}$ as color background. The circles indicate the stable points of nuclear polarization. (a) Overview of the whole region where stable points are expected: Four stable points can be distinguished. (b,c) Close ups around the two stable points with high current (corresponding respectively to points $c$ and $b$ in Fig.\ 4a in the main text). From the background colors we can see that the difference in current is $\sim 15$~\%. To generate these plots we used for both dots $\xi\Gamma_r[E_Z]/\Gamma_s = 0.25$. In the left dot $\Gamma_1/\Gamma_2 = 3.6\cdot 10^{-3}$, $\Gamma_2\tau_n = 9.04$, and $B_0 -\omega = 3.3\, B_1^L$, and in the right dot $\Gamma_1/\Gamma_2 = 16\cdot 10^{-3}$, $\Gamma_2\tau_n = 20.3$, and $B_0 -\omega = 7.1\, B_1^R$.}\label{fig:fig3}
\end{figure}
\begin{figure}[t]
\includegraphics[width=17cm]{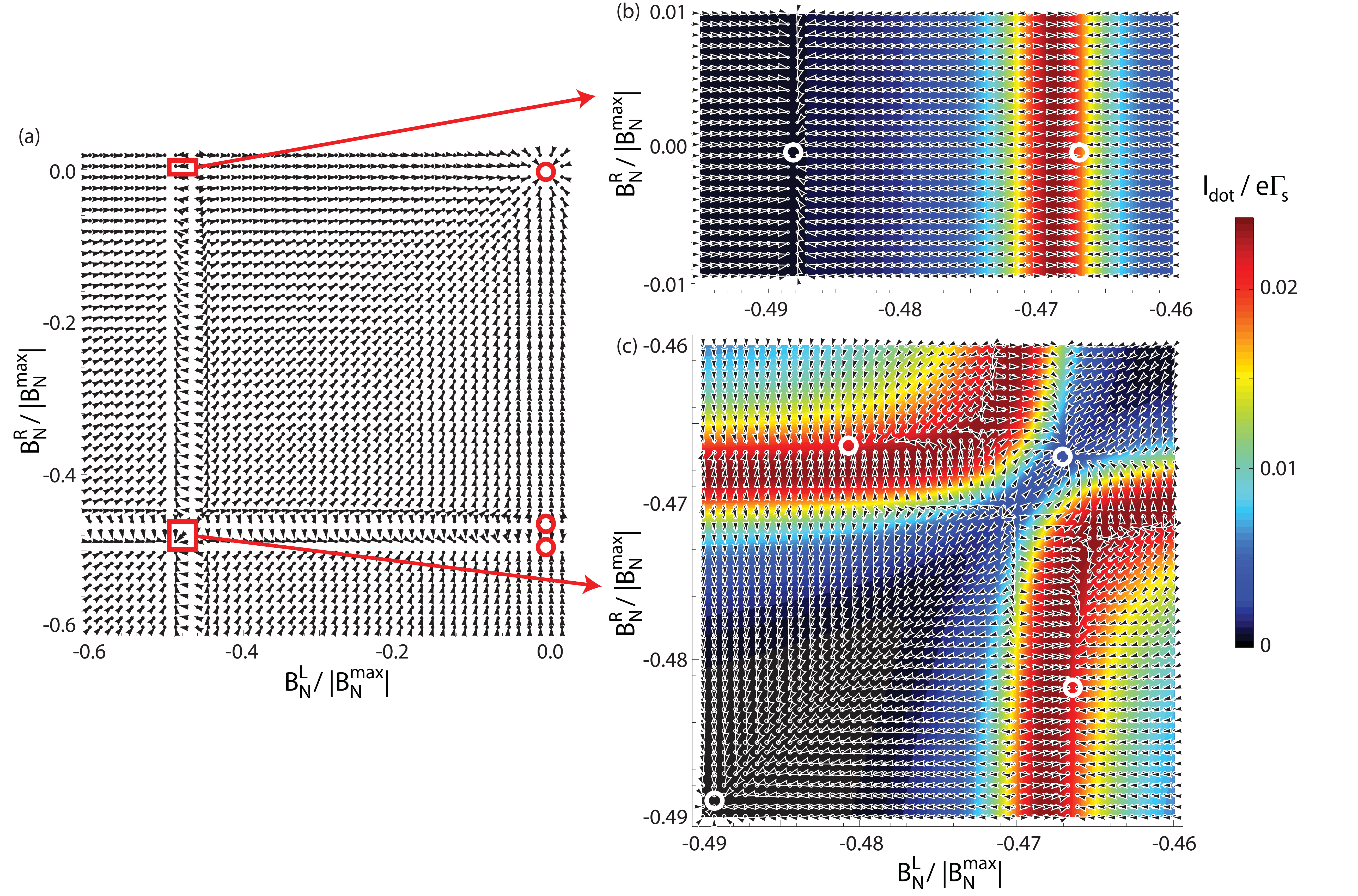}
\caption{In the middle of the hysteresis interval, $B_0 \approx B_\text{res}+ 0.5\ |B_N^\text{max}|$, assuming a symmetric double dot (Fig.\ 4d in the main text). Again we show vector plots of $\{ \dot B_N^L, \dot B_N^R \}$ in the plane $(B_N^L, B_N^R)$. (a) Overview of the whole plane, where the circles indicate the stable points of nuclear polarization. Owing to the double-peak structure of the pumping curve, the left(right) dot has three stable points along the line $B_N^{R(L)} = 0$, i.e.\ where the right(left) dot is unpolarized. (b) Close up of the region where the right dot is unpolarized and the left dot is close to resonance. One of the stable points in this region corresponds to a high current through the system, the other to low current. (c) Close up of the region where both dots are close to resonance. Four additional stable points can be distinguished, two of which correspond to low current and two to high current. To generate these plots, we used $\xi\Gamma_r[E_Z]/\Gamma_s = 0.75$, $\Gamma_1/\Gamma_2 = 21\cdot 10^{-3}$, $\Gamma_2\tau_n = 5$, and $B_0 -\omega = 0.47\, B_N^\text{max}$.}\label{fig:fig2}
\end{figure}\fontdimen2\font=0.8\fontdimen2\font
From the steady state solution of (\ref{eq:fp}) we evaluate the small fluctuations of the nuclear fields around the stable states. For any unpolarized dot we find $\left\langle (\Delta B_N)^2\right\rangle = A_k^2N\equiv \Omega^2$, i.e. the fluctuations are not affected by ESR. If one or both of the dots are polarized, then we can express the resulting nuclear field fluctuations in the polarized dot in terms of the maximally reachable field as $\left\langle (\Delta B_N)^2 \right\rangle \approx (B_1/|B_N^\text{max}|)\Omega^2$, i.e.\ the fluctuations are suppressed by a factor $B_1/|B_N^\text{max}|$.\fontdimen2\font=1.25\fontdimen2{More detailed results}

Here we will present three plots in addition to Fig.\ 4 in the main text. The plots in this section are generated using the same parameters as in Fig.\ 4, but supply some extra details which were omitted from Fig.\ 4 for reasons of clarity: Here we include vector field plots of the time-derivatives $\{ \dot B_N^L, \dot B_N^R \}$ in the plane $(B_N^L, B_N^R)$. Starting from a specific nuclear field configuration $(B_N^L, B_N^R)$, following the arrows shows the evolution in time of the nuclear fields. We added the current through the system $I_\text{dot}$ as color background, this gives a more quantitative picture of the current levels in the different stable points.

As mentioned above, the current through the double dot can be calculated from the quasi-stationary populations $p_n$ as $I_\text{dot} = e\Gamma_s \sum_{n} |\langle S | n \rangle |^2 p_n$, and is a function of $f_{L,R} / \tilde B_{L,R}$ (and therefore of the nuclear polarizations in the two dots), the temperature $\xi$ and the ratio $\Gamma_r[E_Z]/\Gamma_s$. In Fig.\ \ref{fig:fig1s} we plotted $I_\text{dot}$ close to the point where both dots are on resonance, i.e.\ where $f_L=f_R=0$. The function has the structure of two crossing Lorentzians, with a suppression at the resonant point $f_L=f_R=0$. In all current plots in this section, we subtracted the leakage current far away from resonance: It is a measure for the spin relaxation rate $\Gamma_r[E_Z]$.

In Fig.\ \ref{fig:fig3} we replotted Fig.\ 4a from the main text, and added $\{ \dot B_N^L, \dot B_N^R \}$ as vector field and the current $I_\text{dot}$ as color background. In this case $B_0-B_\text{res}\sim B_1$, i.e.\ the detuning of $B_0$ and $\omega$ is relatively small. In the whole plane we distinguish four stable points: two with low and two with high current. As can be seen from the close ups in Figs \ref{fig:fig3}b and c, the two values of high current can differ with ca.\ 15~\%. The asymmetry in $\tilde B_{L,R}$ and $N_{L,R}$ is implemented by using $\Gamma_1^L / \Gamma_1^R = 0.097$ and $\Gamma_2^L / \Gamma_2^R = 0.44$. This corresponds to a difference in a.c.\ magnetic fields $B_1^{L,R}$ of $\sim 50$~\% and a difference in effective numbers of nuclei of $\sim 30$~\%.

In Fig.\ \ref{fig:fig2} we present the same plot as in Fig.\ 4d in the main text, again with the vector field of time evolution and the current added. Here $B_0 \approx B_\text{res}+ 0.5\ |B_N^\text{max}|$, i.e.\ the system is in the middle of the hysteresis interval and we assume a symmetric double dot, i.e.\ equal parameters for both dots. In the whole plane nine stable points can be distinguished. In four of those points the current through the double dot is relatively high. The unpolarized point $B_N^L=B_N^R=0$ is so far away from the other stable states that, as soon as the system switches to the unpolarized state, it will stay there forever.


\begin{thebibliography}{99}
\bibitem{fig2} Another peak is observed at zero field which manifests the mixing of the four (1,1) states by the fluctuating nuclear fields~\cite{frank:nature,frank:science}. An offset of $\sim$ 7 mT of the external magnetic field due to the superconducting magnet is compensated for in both traces.
\end{thebibliography}
\begin{thebibliography}{99}
\bibitem{endnote1} In this expression we assume for simplicity nuclei with spin 1/2. In GaAs $I=3/2$, which gives rise to an extra numerical prefactor in the later results.
\end{thebibliography}
\end{document}